# A LASSO-Inspired Approach for Localizing Power System Infeasibility

Shimiao Li, Amritanshu Pandey, Aayushya Agarwal, Marko Jereminov, Larry Pileggi
Dept. of Electrical and Computer Engineering
Carnegie Mellon University
Pittsburgh, PA, USA

*Abstract*—A method is proposed to identify and localize the cause of network collapse with augmented power flow analysis for a grid model with insufficient resources. Owing to heavy network loading, insufficient generation, component failures and unavoidable disturbances, power grid models can sometimes fail to converge to a feasible solution for a steady-state power flow study. For applications such as system expansion planning, it is desirable to locate the system buses that are contributing to network infeasibilities to facilitate corrective actions. This paper proposes a novel LASSO-inspired regularization of the power flow matrix that enforces sparsity to localize and quantify infeasibilities in the network. One of the examples demonstrates how the proposed method is capable of localizing a source of blackout to a single dominant bus on an 80k+ bus eastern interconnection model.

## I. INTRODUCTION

To ensure grid reliability and security, existing and planned power systems are evaluated on whether they can survive critical contingencies while serving current or forecasted loads. Often, due to severe contingencies, heavy loading, and other limitations, the simulation indicates network collapse, which corresponds to a grid that has likely blacked out [1]. In a traditional power flow study, this collapsed grid state corresponds to no solution, and is characterized by divergence of the simulation [2]. Recently, methods [3],[4] have been developed that instead provide an *infeasible* power flow solution for such collapsed grid states.

These infeasibility-based power-flow methods can converge for a collapsed grid state; however, they do not provide specific cause of power outage, nor do they identify *localized* locations that are disrupting system security and robustness. In most situations, it would be desirable to know the smallest possible set of dominant nodes that are causing system collapse with some quantifiable metric.

Accurate and efficient localization of this dominant set of nodes identifies the deficiency of power (real and reactive) and highlights some critical locations for special attention in the planning process. For instance, consider reactive power planning (RPP) problems [5]-[8] that aim to find the optimal allocation of reactive power support through capacitor banks or FACTS devices such as static VAR compensators (SVC). Such problems correspond to finding the sparsest reactive power compensation vector that satisfies system power balance and operation limits in an optimization-based power flow study. However, convergence of such optimization-based studies becomes more difficult with increased system size and operating limits. Most state-of-the-art placement planning strategies [5]-[8] are only shown to handle small cases with hundreds of buses or less and are known to suffer from lack of robust convergence. Instead, a sparsity enforcement method is preferred. The objective of this method will be to provide a sparse set of nodes, that along with quantified *infeasibility* power, can be added to each of the corresponding set of nodes to make the model feasible.

This paper proposes a novel method to localize infeasibility within power grid models that have no solution. Unlike existing formulations [3],[4] that distribute system infeasibilities across all buses in the system via minimization of the sum of square of their values (L2-norm), the proposed approach localizes the infeasibility to a few system buses. The approach is inspired by LASSO [9],[10], a method that is used to enforce sparsity in feature selection of a model by L1-regularization. In our approach, the sparsity is enforced on the infeasibility solution vector to obtain infeasibilities at only the dominant nodes in the system. This paper defines a new approach of enforcing sparsity such that the infeasibilities corresponding to geographically localized buses are correlated through a *bus-wise sparsity enforcer*.

Mathematically, this problem is formulated as a non-convex optimization problem and is implemented based on an equivalent circuit formulation (ECF) that has been demonstrated to ensure convergence for power flow analysis and associated optimization applications (to a local optima), such as that for identifying infeasibility.

Since the power system is highly nonlinear, the major objective of our method can be summarized as follows:
i. To reach a sufficiently **sparse** solution; i.e., supply additional real or reactive power at minimum number of possible locations for an infeasible network model.
ii. To facilitate robust convergence for **large-scale** systems.

Section II gives a background overview of the ECF [3],[11],[12] approach. Section III lists a series of power flow related formulations to illustrate the quantification of system infeasibility. Section IV presents the exploration of a sparsity enforcing mechanism, insight into problems with L1-regularization based optimization problems, the proposed infeasibility localization algorithm, and application to large scale systems with reliable convergence. Section V presents some experimental results on several large cases that include the U.S. Eastern Interconnection sized 80k+ buses network, followed by our conclusions in section VI.

## II. BACKGROUND

### A. Notation

Table 1 shows the symbols used in this paper.

TABLE 1. SYMBOLS AND DEFINITIONS

| Symbol | Interpretation |
| --- | --- |
| $V^R, V^I, I^R, I^I$ | Real/imaginary voltage/current; |
| $|V|, \theta, P, Q$ | Voltage magnitude/angle; active/reactive power |
| $X$ | Power flow solution vector, $X = [V^R, V^I]$ |
| $I_{f,i}^{R/I}$ | Real or imaginary infeasibility current at bus #i |
| $I_f$ | Infeasibility current vector |
| $I(x) = 0$ | KCL equations at all buses |
| $k$ | Sparse goal: to localize infeasibility to $k$ locations |

### B. Equivalent Circuit Formulation (ECF)

A circuit-theoretic formulation for power grid analysis was developed in [3],[11],[12]. Instead of describing components with 'PQV' parameters, the ECF framework models each component within the power grid as an electrical circuit element characterized by its I-V relationship. For computational analyticity, the complex relationships are split into real and imaginary sub-circuits whose nodes corresponds to power system buses [3][11][12]. Table 2 shows a simple comparison between the traditional PQV formulation and ECF.

TABLE 2. COMPARISON BETWEEN FORMULATIONS

| Property | Comparison | |
| --- | --- | --- |
| | PQV formulation | ECF (I-V formulation) |
| Coordinate | Polar | Rectangle |
| State variables | $|V|, \theta$ | $V^R, V^I, Q$ |
| Network balance | Zero power mismatch | Zero current mismatch |
| Governing equations | Power balance eqns | KCL eqns |
| Network constraints | Non-linear | Linear |
| Loads and generators | Linear | Non-Linear |

Under this ECF framework, all branches, e.g. transmission lines, transformers, and shunts, are linear components, due to the intrinsic linearity of their I-V relationships. Other components such as generators and loads have nonlinear I-V models. The system balance is expressed by a set of nonlinear KCL equations:

$$I(x) = 0 \quad (1)$$

These equations can be iteratively linearized and solved via Newton Raphson (NR) using bus voltage variables. Importantly, since ECF enables power systems of any size to be efficiently simulated as an equivalent circuit, numerous convergence techniques that were developed for circuit simulation (e.g. SPICE [15]) can be applied.

## III. POWER FLOW RELATED PROBLEMS

Next, we present a series of power flow formulations that introduce our proposed method for quantifying and localizing infeasibility.

### A. (Problem Approach 1) Traditional power flow

Traditional power flow outputs the bus voltage solution by iteratively solving the nonlinear network balance equations in (1). For a feasible network model, the traditional power flow converges to a feasible solution; however, if no feasible solution exists for the network, the methodology diverges resulting in no useful solution.

### B. (Problem Approach 2) Infeasibility-quantified power flow

To avoid divergence without providing information, an approach to capture the infeasible-quantified power flow was developed. To effectively distinguish between hard-to-solve network case from infeasible network case, extensions have been proposed and developed for both the ECF-based [3][11] and traditional [4] methods to quantify the potential infeasibility within the grid. The ECF-based approach [11] introduces '**infeasibility current**' $I_{f,i}^R, I_{f,i}^I$ at each bus #i. These values represent compensation terms that capture how much additional real/imaginary current flow is needed at each bus to make the network balance conditions hold. The infeasibility-quantified power flow study is formulated as a non-convex optimization problem:

$$\min_{X, I_f} \frac{1}{2} \left|\left| I_f \right|\right|_2^2 \quad (2)$$
$$s.t. \text{ system balance equations}: I(X) + I_f = 0$$

where $I_f$ is infeasibility current vector that contains $I_{f,i}^{R/I}$ for $\forall i$.

This formulation can clearly identify a feasible case from an infeasible one:
- Convergence with zero infeasibility currents everywhere denotes system balance.
- Convergence with nonzero $I_{f,i}^{R/I}$ denotes an infeasible system with specific power flow deficiency at each bus $i$.
- Divergence is totally attributed to insufficient convergence robustness of the algorithm.

### C. (Problem Approach 3) Proposed: Infeasibility-localized power flow by L1-regularization

Mathematically, solving network balance equations with inclusion of $I_f$ is an under-determined problem that has infinite solutions. Problem Approach 2 outputs an optimal solution with minimal L2 norm, however, this solution is not designed to be sparse. Therefore, the quantities and locations of the nonzero infeasibility currents is not necessarily an informative identifier of the dominant sources of the modeled outage. Especially in large scale systems, we can have numerous comparative infeasibility currents across the grid. In reality, the most useful solution for expansion planning and corrective action is the not the one with the smallest L2 norm and infeasible current values at multiple sources, but rather a solution that has non-zero infeasible currents in the least number of locations.

To address this, a classical method for enforcing sparsity can be applied using the L1-norm in the objective function:

$$\min_{X, I_f} \frac{1}{2} \left\| I_f \right\|_2^2 + c \left\| I_f \right\|_1 \quad (3)$$
$$s.t. \, I(X) + I_f = 0$$

However, this formulation neglects the correlation of real and imaginary counterparts during sparsity enforcement, while in reality, the nonzero $I_{f,i}^R, I_{f,i}^I$ at the same bus are often coupled terms emerging concurrently. Moreover, the desired sparseness requires higher values assigned to regularization parameter $c$, making it difficult for the algorithm to converge. More detailed explanations for the convergence difficulty are explored in section IV.

## IV. PROPOSED METHOD

To present our new approach, we first exploit some physical intuition of the system and develop a general methodology (within sub-section A) that enforces a sparse system solution using problem approach 3. Also, in this sub-section, we discuss challenges of this approach from the convergence perspective. Next (within sub-section B), based on the observed mechanism, we propose our basic idea of bus-wise sparsity enforcer, a novel regularization term to eliminate the aforementioned limitations.

### A. An insight into sparsity inforcement by L1 regularization

In the L1-regularization-based sparse enforcement method (**problem approach 3**) developed above, the inclusion of L1-norm leaves an undifferentiable objection function. To tackle this, we introduce slack variable $t$ and convert the problem into the following constrained optimization form [13]:

$$\min_{X, I_f} \frac{1}{2} \left\| I_f \right\|_2^2 + c \cdot t \quad (4)$$
$$s.t. \quad I(X) + I_f = 0 \quad (5)$$
$$I_f \leq t \quad (6)$$
$$-I_f \leq t \quad (7)$$

where the slack variable vector $t$ represents the upper bound on the infeasibility currents vector $I_f$. Each $I_{f,i}^{R/I}$ corresponds to an upper-bound $t_{f,i}^{R/I}$ such that $|I_{f,i}^{R/I}| \leq t_{f,i}^{R/I}$, as in (6)-(7). We can write its Lagrangian function as:

$$L(X, I_f, t, \lambda, u_{U/L}) = \frac{1}{2} \left\| I_f \right\|_2^2 + ct + \lambda^T (I(X) + I_f) \quad (8)$$
$$+ \mu_U (I_f - t) + \mu_L (-I_f - t)$$

The perturbed KKT conditions of this problem are:

$$I_f^* = -\lambda^* - u_U^* + u_L^* \quad (9)$$
$$u_U^* + u_L^* = c \cdot 1 \quad (10)$$
$$u_U^* (I_f^* - t^*) = -\epsilon \quad (upper \, bound) \quad (11)$$
$$u_L^* (-I_f^* - t^*) = -\epsilon \quad (lower \, bound) \quad (12)$$

By further manipulation based on properties of Lagrangian multiplier, the primal-dual pair $(I_f, \lambda)$ should satisfy:

$$|I_f^*| = |\lambda^*| - c \quad (13)$$

This primal-dual relationship can be clearly illustrated by Figure 1, and inspires us to attach intuitive **physical meanings**:

- Bus-wise Lagrangian multiplier $\lambda_i^{R/I}$ is a source of additional current flow into the network
- Scalar $c$ is a threshold such that any infeasibility quantities $\lambda_i^{R/I}$ below threshold are blocked out and only those above this threshold become the $I_{f,i}^{R/I}$ 'flow' into the system.

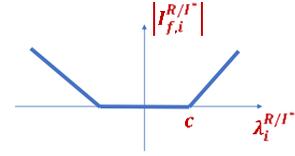

Figure 1. Relationship between $I_f^*$ and c: a blocking effect

This reveals a simple **mechanism** through which the threshold $c$ encourages a sparse solution by confining most $I_f$ to near zero value. Whenever threshold $c$ is added, **the blocking effect** reduces the number of non-zero infeasibility sources in the network. As the threshold $c$ is increased, the number of non-zero infeasibility sources decrease, and any remaining non-zero infeasibility sources adjust their value to make the network feasible. Therefore, with a high enough threshold value, only a few sources turn out to be above threshold and appear as nonzero elements in $I_f$.

In summary, our approach utilizes that: **raising the value of c encourages more near-zero $I_{f,i}^{R/I}$ elements by making the threshold hard-to-pass.**

However, there exists serious convergence problem with a single scalar $c$ as tuning parameter for regularization. This challenge can be characterized by *an unwanted trade-off & inflexibility*. Let us illustrate this further.

With $t$ representing the upper bound of $I_f$, if there exists nonzero infeasibility (e.g. $I_{f,i}^R > 0$) at bus $i$, due to the minimization of $t$ in the objective function, the upper bound tends to be very tight (i.e., $t_i^R - I_{f,i}^R \to 0$). Hence, if we utilize a single large scalar $c$ value to achieve a sufficiently sparse solution, the tightness property of the algorithm results in convergence difficulties due to the steep and highly non-linear regions of the complementarity slackness conditions given by (11)-(12).

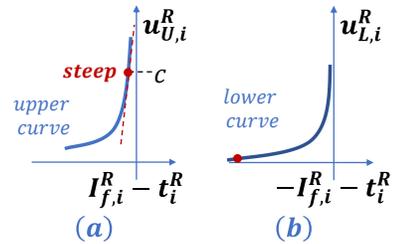

Figure 2. High c value causes steep convergence region on the complementary slackness curve: (a)upper bound curve $u_{U,i}^R (I_{f,i}^R - t_i^R) = -\epsilon$, (b)lower bound curve $u_{L,i}^R (-I_{f,i}^R - t_i^R) = -\epsilon$. When $I_{f,i}^R > 0$, we have $u_{U,i}^R \to c, u_{L,i}^R \to 0$. $(u_{U,i}^R, I_{f,i}^R, t_i^R)$ converge on a difficult region of the upper curve

This problem can be illustrated in Figure 2. As the number of infeasible buses increases, numerous buses encounter difficult steep regions of this kind, making it difficult for the algorithm to converge.

Thus, the selection of the value of the $c$ parameter is a

**trade-off** between sufficient sparsity and robust convergence, both of which are essential for meeting our eventual goal. With $c$ being a single scalar value, there is little freedom for us to manipulate its value and achieve the desired performance.

*B. Proposed method: Bus-wise sparsity enforcer*

To address the aforementioned challenges, we propose a new method that defines threshold $c_i$ for each bus $i$. This $c_i$ parameter is a *bus-wise sparsity enforcer* such that, according to the thresholding effect, raising $c_i$ encourages a zero $I_{f,i}^{R/I}$ value at bus $i$ in the solution. The infeasibility localization problem can now be reformulated as problem 4.

**(Problem Approach 4)** *Infeasibility-localized power flow by bus-wise sparsity enforcer*

$$\min_{X, I_f} \frac{1}{2} \left\| I_f \right\|_2^2 + \sum_i c_i (|I_{f,i}^R| + |I_{f,i}^I|) \quad (14)$$
$$s.t. \quad I(X) + I_f = 0$$

In this approach, we convert a single scalar $c$ to a vector of bus-wise sparsity enforcers. Then, to determine the values of $c_i$, we use the following assumptions that are based on the grid physics:

- Uneven distribution of infeasibility sources: system infeasibility is caused by and can be characterized by failures on isolated locations, rather than outages of equal seriousness at each bus.
- There is a high probability that the dominant sources (locations) of failure in the system are reflected by the nodes with highest magnitude of $I_{f,i}^{I/R}$ in the simulation.

Based on these assumptions, we can simply make flexible adjustments to $c_i$ at each bus, according to the qualitative classification of bus-wise infeasibilities, as shown in Algorithm 1. For simplification and efficiency, we simply classify all buses into '**major**' and '**minor**' categories, according to their infeasibility current magnitude and the sparsity goal. For buses in the 'major' group, i.e. with high infeasibility quantities ($|I_{f,i}^{I/R}| \gg 0$), we assume that they are very likely the dominant sources of failure and assign a low value $c_L$. This encourages nonzero infeasibility current on those locations. For buses in the minor group, we assign a higher threshold $c_H$ such that we can force their infeasibility values to zero or near zero values.

---
**Algorithm 1: Bus-wise Sparse Enforcer Assignment**
---
**Input**: sparse goal $k$, threshold $(c_H, c_L)$, infeasibility current $I_f$
**Output**: updated bus-wise sparsity enforcer $c_i, i = 1,2,\dots, n_{bus}$
1. Calculate infeasibility current magnitude at all buses $I_{mag}$
2. Classify bus category:
   'major' buses index: $id_{major} = argsort(I_{mag}, k)$
   'minor' buses index: $id_{minor}$ contains the remaining buses
3. Assign $c_i$
   $c(id_{major}) = c_L$, $c(id_{minor}) = c_H$
---

Our infeasibility localization method is summarized in Algorithm 2, where $k$ defines the number of locations where non-zero value of infeasibility sources might be allowed. From another intuitive viewpoint, this method unevenly penalizes infeasibility values at different buses. By assigning high $c_i$ values to certain buses, we deliberately attach high penalty to infeasibility currents in those buses, thereby forcing the infeasibility currents to make the network feasible from other sets of buses with low $c_i$ attached to them. More importantly, this $(c_H, c_L)$ configuration removes the need for high values of parameter $c_H$, as sparsity is dependent on the ratio of $c_H$ and $c_L$, not the absolute value of the threshold. Simple principles for selecting $(c_H, c_L)$ are:

- $c_H$ is chosen to be sufficiently larger than $\lambda_i^{R/I}$ such that 'minor' infeasibility sources result in zero or near zero values. This enables sufficiently sparse solution with small enough $(c_H, c_L)$ values, thereby avoiding convergence difficulties.
- $c_L$ is chosen to be sufficiently lower than $c_H$ such that the threshold is 'easy-to-pass' for both $\lambda_i^R$ and $\lambda_i^I$, making nonzero $I_{f,i}^R$, $I_{f,i}^I$ coexist at infeasible buses. This is a necessary condition for practical applications. Due to the nature of the power flow equations, grid devices provide both real and imaginary currents. Therefore, for any corrective actions, it is preferable to achieve sparse solutions that have infeasibilities localized to the fewest number of buses, rather than fewest number of nonzero $I_{f,i}^{R/I}$.

Additionally, if the k-sparse goal is not practical, $I_{f,i}^{R/I}$ in the 'minor' group leaves room for infeasible sources on more than $k$ locations, and the final solution can be $(k + m)$-sparse.

---
**Algorithm 2: Infeasibility Localization with k-sparse goal**
---
**Input**: testcase, initial guess $(X_0, I_{f0}, \lambda_0)$, sparse goal $k$, threshold $(c_H, c_L)$
**Output**: a sparse $I_f$ vector
1. Initialize $t, u_U, u_L$
2. **Bus-wise sparse enforcer assignment**
3. Infeasibility-localized power flow by bus-wise sparsity enforcer (**Problem4**)
---

*C. Extension to large-scale systems*

For a practical large-scale power system, we do not have accurate knowledge in advance about the severity of the system collapse, and therefore, it is hard to define a reasonable guess of the *k*-sparse goal. Importantly, since infinite possible combinations of $\{I_{f,i}^{R/I}\}$ can make the system network balance equations correspond to a feasible network, the 'major' locations in a dense solution are likely to be the dominant sources with a high probability; however, we must note that this is not always true.

---
**Algorithm 3: Infeasibility Localization for large-scale systems**
---
**Input:** testcase, shrinkage rate $r$
**Output**: a sparse $I_f$ vector
1. Initialize $(X_0, I_{f0}, \lambda_0)$ by infeasibility-quantified power flow (**Problem2**)
2. Initialize $(c_H, c_L)$
3. Initialize sparsity goal: $k = n_{bus} * r$
3. **while** not sparse enough, **do**
   **Bus-wise sparse enforcer assignment**
   **Infeasibility Localization with k-sparse goal**
   Check current solution sparisty $k$
   Update sparsity goal: $k = k * r$
   (*Optional*) adjust $(c_H, c_L)$ if needed
   (*Optional*) adjust shrinkage rate $r$ if needed
---

With these considerations we extend our method to large-scale networks by iteratively adjusting sparse enforcers and gradually reaching sparser solution from denser ones. For robust convergence, we start from a dense solution from (**problem approach2**) robust regular power flow [3][14] with quantified infeasibility in all locations and gradually update the k-sparse goal by some shrinkage rate. This is equivalent to splitting the original problem into a series of subproblems, where each subproblem uses a solution from the previous one as its initial guess, and easily reaches its optimal solution within a few iterations. Our method is shown in Algorithm 3.

## V. EXPERIMENTS

This paper conducts experiments to prove the efficacy and scalability of our proposed method. To create an infeasible scenario (past the nose curve) on these cases, we increase their loading factor $\alpha$. And parameters of our proposed method are set to default values $c_H = 10, c_L = 0.1, r = 0.75$.

We first test standard CASE14 which is infeasible at $\alpha = 4.5$. Table 3 presents infeasibility, quantified first by infeasible power flow [3] (**Problem2**) and sparsity enforcement using L1-regularization (**Problem3**) and our proposed method (**Problem4**). Comparison shows that our method reaches 1-sparse solution and localizes infeasibility to bus#14, whereas the standard infeasibility approach [3] localizes infeasibility to almost all buses making the approach impractical for expansion planning or applying corrective action.

TABLE 3. METHOD COMPARISON RESULTS ON CASE14

| Bus ID | Infeasibility current magnitude solution $|I|$ | | |
|---|---|---|---|
| | Non-sparse [3] | L1-regularization | Proposed method |
| 1 | 0 | 0 | 0 |
| 2 | 0.00858402 | 0 | 0 |
| 3 | 0.0561223 | 0 | 0 |
| 4 | 0.05097014 | 0 | 0 |
| 5 | 0.04278203 | 0 | 0 |
| 6 | 0.08877886 | 0.16111856 | 0 |
| 7 | 0.07740694 | 0 | 0 |
| 8 | 0.09593462 | 0.33915759 | 0 |
| 9 | 0.08860328 | 0 | 0 |
| 10 | 0.09134275 | 0 | 0 |
| 11 | 0.08889756 | 0 | 0 |
| 12 | 0.09065051 | 0.1244972 | 0 |
| 13 | 0.09368859 | 0.27381069 | 0 |
| 14 | 0.10908567 | 0.1824952 | 0.80006182 |

Next, we test 5 large system cases. Table 4 shows our method efficiently localizes system infeasibility to sparse distributions.

TABLE 4. RESULTS

| Case Name | α | k-sparse | Dominant infeasible buses name |
|---|---|---|---|
| MMWG80K | 1.07 | 1 | '155753' |
| ACTIVSg25K | 1.8 | 42 | Not listed here |
| CASE9241pegase | 1.15 | 1 | '2159' |
| CASE6515rte | 1.15 | 2 | '3576,' 4356' |
| CASE6468rte | 1.29 | 1 | '3718' |

## VI. CONCLUSION

This paper presents a novel approach to localize the source of infeasibility in a grid network model. This is mathematically equivalent to finding a sparser solution to an underdetermined nonlinear system. With L1-regularization suffering from limited solution sparsity due to the unwanted trade-off and lack of flexibility in the parameter adjustment, we propose a new method based on the physics-based models and mechanisms corresponding to bus-sparsity enforcement of the L1-norm. The primary contributions of our approach are:

- **Definition of bus-wise sparse enforcer** $c_i$ to replace the scalar parameter in L1-norm.
- **Creation of a new regularization term** with uneven blocking effect (penalization) on each bus.
- **Manipulation of sparsity** by adjusting enforcers, based on the observed sparse mechanism.

ACKNOWLEDGMENT

This work was supported in part by the Defense Advanced Research Projects Agency (DARPA) under award no. FA8750-17-1-0059 for RADICS program, and the National Science Foundation (NSF) under contract no. ECCS-1800812.